\documentclass[reprint,aps,prb,groupedaddress,floatfix]{revtex4-2}
\usepackage{graphicx}
\usepackage{bm}
\usepackage{dcolumn}

\input{epsf}

\begin{document}
\title{Magnetoresistance of edge states of a two-dimensional topological insulator}

\author{Leonid Braginsky}
\email{brag@isp.nsc.ru}
\affiliation{\it Institute of
Semiconductor Physics, 630090 Novosibirsk,
Russia}
\affiliation{\it Novosibirsk State University, 630090 Novosibirsk,
Russia}

\author{M. V. Entin}
\email{entin@isp.nsc.ru}
\affiliation{\it Institute of
Semiconductor Physics, 630090 Novosibirsk,
Russia}
\affiliation{\it Novosibirsk State University, 630090 Novosibirsk,
Russia}

\date{\today}
\begin{abstract}
The theory of magnetoresistance of the edge state of a two-dimensional topological insulator is developed. The magnetic field violates the  time reversal invariance. Magnetoresistance arises due to  the energy gap opened by a magnetic field parallel to the sample surface.   The combined action of impurities and the magnetic field causes backscattering of edge electrons. Although impurities are necessary for scattering, a sufficiently strong interaction with impurities leads to the suppression of backscattering.
\end{abstract}
\maketitle




\section{Introduction}
 The edge states of 2D topological insulator is one of most inspiring problem of the modern solid state physics. It was shown that these states possess so called topological protection, preventing the electron from backscattering. At a moment there is a large and rapidly growing number of publications  on this topic, (see, i.g.,  \cite{01,02,03,04}).
The study of the conductance of the edge states of a two-dimensional topological insulator assumes their topological protection, this makes backscattering prohibited or it is extremely weak. Therefore, the one-dimensional states are collisionless and the conductance of the electrons on them is $ e ^ 2 / h $.

There were some attempts to implement mechanisms violating the time-reversibility and, therefore, causing backscattering. In particular, Ref.~\cite{glazman1,glazman2} ascribe backscattering to the transitions between edges with opposite direction of travel  of the same-spin electrons. The transition between opposite edges of a TI strip were considered also in Ref.~\cite{magent}.

The present paper was stimulated by the experimental finding  \cite{Piatrusha2019} of the strong magnetoresistence of the edge state electron conductance. Important experimental fact that has to be explained is the presence of magnetoresistance, as well as its gigantic value and fluctuations with the Fermi level. What is unusual, is the sensitivity of 1D edge states to the magnetic field, which is absent in other 1D systems. The results of \cite{Piatrusha2019} can not be explained by means of interedge transitions \cite{magent} due to large width of the TI strip in the experimental conditions.

The model is based on the one-dimensional Hamiltonian

\begin{eqnarray}\label{1}
H=\left(
    \begin{array}{cc}
      vp+V(x) & \Delta_1 \\
      \Delta_1^* & -vp+V(x) \\
    \end{array}
  \right).
\end{eqnarray}
Here $p$ and $v$ are the electron momentum and velocity, and $V(x)$ is the impurity potential. The Hamiltonian (\ref{1}) is the simplest that violates the time reversal invariance in the magnetic field and leads to the gap $2|\Delta_1|$  at $p=0$.

 The gap in the electron spectrum of the 2D topological insulator arises in the magnetic field   ${\bf B}$ directed along the specimen plane.\cite{Durnev2016}
 \[
 \Delta_1=\mu_B(g_{xx}B_x+ig_{yy}B_y)
 \]
 for the (0,0,1) orientation.
 Here $\mu_B$ is the Bohr magneton, and $g_{xx},~g_{yy}$ are the  $g$-factor components in the specimen plane.

  The gap is
  \[
 2|\Delta_1|=\mu_B\sqrt{g_{xx}^2B_x^2+g_{yy}^2B_y^2}.
 \]

At a weak magnetic field the magnetoresistance is due to the backscattering and localization of previously delocalized states by the magnetic field. Indeed, the presence of off-diagonal terms in the Hamiltonian leads to backscattering of electrons. The potential itself does not cause transitions between the states. The scattering occurs due to the off-diagonal part $ H_ {nd} $ of the Hamiltonian (\ref{1}).

Two approaches can be utilized. First is the classical approach. In the absence of localization the 1D conductivity is of the Drude form:
\begin{equation}\label{51}
\sigma= \frac{e^2\tau v}{\pi}.
\end{equation}
Since the electron velocity in the linear spectrum is constant, the conductivity is determined by the backscattering time $ \tau $. The classical approach is valid in the absence the phase coherence at a finite temperature (similar problem has been considered in connection to the TI strip \cite {magent}).  This means that during the forward and backward travel at the distance $ v \tau $ the phase of the electron changes by $ \pi $ due to, e. g., interaction with phonons.

Second, the coherence of the electron states results in their localization in the infinite volume and, therefore, exponential size dependence of the conductivity at a zero temperature. In this case the 1D conductivity is not established. Instead, the conductance $G$ of the system drops exponentially with the system length $L$: $\log G\propto -L/v\tau$.

In this paper we find the backscattering time $ \tau $.

\subsection*{Magnetoinduced backscattering}
The selfwave functions of the  electrons of the Hamiltonian (\ref{1}) with energies  $\epsilon_{\pm}=\pm vp$ at $B=0$  are
  \begin{eqnarray*}\label{2}
  &&\psi_+=\frac{1}{\sqrt{L}}\left(
         \begin{array}{c}
           1 \\
           0 \\
         \end{array}
       \right)e^{ipx+i\int dx V(x)/v}\\
  &&\psi_-=\frac{1}{\sqrt{L}}\left(
         \begin{array}{c}
           0 \\
           1 \\
         \end{array}
       \right)e^{ipx-i\int dx V(x)/v}.\nonumber
  \end{eqnarray*}
The amplitude of the backscattering is determined by the first order perturbation on the non-diagonal part of the Hamiltonian $H_{nd}\propto\Delta_1$, i.e., the matrix element between the $|\psi_{-p,-\sigma}\rangle$ and $|\psi_{p,\sigma}\rangle$ states of the same energy $\epsilon_{p,\sigma}$:
 \begin{eqnarray*}\label{333}
&&A=\langle\psi_{-p,-}|H_{nd}|\psi_{p,+}\rangle\\&&=\frac{\Delta_1}{L}\int_{-L}^{L} \exp{\left(2ipx+\frac{2i}{v}\int V(x)\,dx\right)\,dx}.
\nonumber
\end{eqnarray*}

\section{Weak impurity potential. The perturbation theory}
In the first order with regard to the potential  $V(x)$ we find

\begin{eqnarray*}\label{4}
&&A=2i\langle\psi_{-p,-}|H_{nd}|\psi_{p,+}\rangle=\frac{2i \Delta_1}{L pv} \tilde{V}(2p),
\nonumber\\
&&\tilde{V}(p)=\int_{-\infty}^\infty V(x) e^{ipx}dx.
\end{eqnarray*}

 Consider the electron scattering at the impurity potential $V({\bf r})=\sum_n u({\bf r}-{\bf r}_n)$, where $r_n$ is the position of  n-th impurity and $u({\bf r}-{\bf r}_n)$ is its potential.
The probability of  backscattering has to be averaged over the random impurity positions in 2D layer, so that
 \begin{equation}\label{201}
   1/\tau=\frac{L}{ v} \langle|A|^2\rangle,
\end{equation}
The 2D Fourier transform of the Coulomb impurity is
\[
   u(q)=\frac{2\pi e^2}{\kappa q},
 \]
therefore
\[
  \langle |V(q_x)|^2\rangle=n_i\int\frac{dq_y}{2\pi}u^2(q)=2\pi^2 n_i \frac{e^4}{q_x\kappa^2}, ~~~q^2=q_x^2+q_y^2.
\]
Here $n_i$ is the impurity density. For the backscattering probability we find
\begin{equation}\label{207}
   1/\tau_0=\frac{8\pi^2 e^4n_i |\Delta_1|^2}{\hbar \kappa^2 (vp_F)^3}=\frac{8  e^4n_i|\Delta_1|^2}{\hbar^4\kappa^2v^3 \pi n_e^3}.
\end{equation}
Here  $n_e$  is the 1D electron density.

The probability (\ref{207})  essentially increases at a small electron density. Therefore,  screening of the Coulomb potential by the field electrode has to be taken into account. Then

\begin{equation}\label{23}
   u(q)=\frac{2\pi e^2}{\kappa q}(1-e^{-2qd}).
\end{equation}
Here $d$ is the distance to the field electrode. The last factor in Eq.~(\ref{23}) restricts the impurity matrix elements at small $q$.
For $1/\tau$ value we obtain

\begin{eqnarray}\label{208}
  &&\frac{1}{\tau}= \frac{8\pi^2 e^4n_i |\Delta_1|^2}{\hbar \kappa^2 (vp_F)^3}\phi(p_Fd),\\
  &&\mbox{   where}~~~\phi(y)=\frac{2}{\pi}\int_0^\infty \frac{(1-e^{-4y \cosh t})^2}{\cosh t}dt.\nonumber
\end{eqnarray}
    $\phi(y)\approx 16y\ln2/\pi$ at $y\ll1$  and   $\phi(y)\to 1$ at $y\to \infty$.
Thus, screening reduces the divergence of $1/\tau$ value at small electron density: $\tau\propto n_e^2$, instead of  $\tau\propto n_e^3$ for $n_e\to 0$.

It should be noted that the obtained in this section expressions hold only  for the small impurity potentials. This means applicability of the perturbation theory for this potential as well as for the gap value.

  \section{Strong impurity potential}
  The conductivity of electronic gas with quadratic spectrum is too complicated, if it considered outside the frames of the perturbation theory. This is not the case for the electronic 1D gas with the linear spectrum, because this problem has an exact solution.\cite{bragent} This allows us to obtain the result at   $n_e\to 0$.

  Consider the problem of small gap, but arbitrary impurity potential. Assume the random impurity potential obeys the Gaussian distribution. This is correct, if  $|u({\bm r-r_n}|\ll vp_F$. Then in Gauss approximation from Eq.~(\ref{333})  we obtain:

 \begin{eqnarray*}\label{203} \nonumber
 && \langle|A|^2 \rangle=  L^{-2}|\Delta_1|^2\int\int \exp\Bigg[2ip(x-x')-\\&&\frac{4}{v^2}\int\int_{x'}^xdx_1dx_2W(x_1-x_2)\Bigg] dxdx',
\end{eqnarray*}
where
\[W(x-x')\equiv \int_{-\infty}^\infty dk(2\pi)^{-1}e^{-ik(x-x')}\tilde{W}(k)=\langle V(x)V(x')\rangle
\]
is the potential correlation function. Its Fourier transform is
 \[
 \tilde{W}(k)=\frac{4\pi^2 e^4}{\kappa^2}n_i\frac{\phi(kd/2)}{k}.
 \]
 Then using Eq.~(\ref{201}) we obtain

\begin{eqnarray}\label{55}
&&\frac{1}{\tau}=2(d/v)|\Delta_1|^2\\
&&\times\int_{-\infty}^{\infty} dz \cos(s z) \exp\left(-\beta\int \frac{dy}{y^3}\phi(y/2)\sin^2(yz)\right),\nonumber \\
&&s=4p_Fd,\beta=\frac{2\pi e^4}{\kappa^2v^2}n_id^2.\nonumber
\end{eqnarray}

\begin{figure}
\leavevmode\centering{\epsfxsize=0.5\textwidth\epsfbox{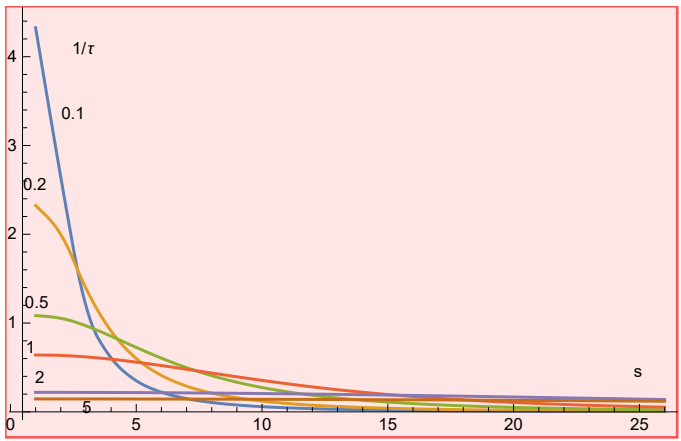}}
\caption{(Coloronline) Dependence of $1/\tau$ on the parameter $s$. Values of $\beta $ are shown on the curves.}\label{Fig2}
\end{figure}

The behavior of $1/\tau$ {\it vs} $\beta$   and $s$  is presented in Fig.~\ref{Fig2}.
The dependence on $\beta$ is linear at small $\beta$  [or at small
impurity density, in agreement with Eqs.~(\ref{207}, \ref{208})]] followed by an
exponential decrease.
The latter is due to the random potential
that results in fluctuating increase of the local Fermi
momentum  reducing the scattering.

\subsection*{Conductance of the finit-size electron system at zero temperature}
 Without backscattering   the zero temperature conductance of the edge state $G$ is equal to the conductance quantum $G_0$ for any edge length. Presence of the backscattering  leads to the localization of the edge states and, therefore, to the exponential drop of the conductance, if the edge length exceeds the backscattering length:  $\ln{G/G_0}\propto-L/v\tau$. Or $\ln{G/G_0}\propto -B^2$  in agreement with the obtained behavior of  $\tau$.

\section*{Discussions}

To estimate the edge state conductivity at a low temperature, consider the 1D electron gas with the linear spectrum. The electron-phonon interaction affects  the phase of the electron wave function, however, this phase is not important for the scattering time. This means that the conductivity obeys the Drude expression Eq.~(\ref{51}) with the relaxation time found in the present paper.

Thus, we found that  a magnetic field parallel to the plane of the sample results in elastic backscattering of  the edge states electrons with the linear spectrum. This scattering leads to localization of states in an infinite-size system and, therefore, to finite conductivity at a non-zero temperature. The localization is due to magnetic field. The transition between the ballistic and localization behavior of conductance occurs under magnetic field in a finite-size sample. This transition depends on the relation between the magnetic localization length and sample length. It is shown that the probability of backscattering increases with the impurity concentration at a low concentration followed by  decrease at a high concentration of impurities.

Note also that our results correct, if only the Fermi level is far apart from the gap $|E_F|\gg|\Delta_1|$. This is necessary for the expansion over the magnetic field holds.

Another note concerns non-magnetic mechanisms of backscattering. The paper \cite{mahent} showed that in a TI with smooth edge the overlapping of linear topology-protected edge states with the Dirac gapped branches leads to the elastic backscattering. In this case the magnetoinduced backscattering complements the latter in the energy domains where they coexist.

The work is supported by the RFBR grant No.20-02-00622.

\end{document}